\documentclass[twocolumn,english,aps,prx]{revtex4}
\usepackage[T1]{fontenc}
\usepackage[latin9]{inputenc}
\usepackage{amsmath}
\usepackage{graphicx}

\makeatletter
\@ifundefined{textcolor}{}
{%
 \definecolor{BLACK}{gray}{0}
 \definecolor{WHITE}{gray}{1}
 \definecolor{RED}{rgb}{1,0,0}
 \definecolor{GREEN}{rgb}{0,1,0}
 \definecolor{BLUE}{rgb}{0,0,1}
 \definecolor{CYAN}{cmyk}{1,0,0,0}
 \definecolor{MAGENTA}{cmyk}{0,1,0,0}
 \definecolor{YELLOW}{cmyk}{0,0,1,0}
 }
\makeatother

\usepackage{babel}
\begin{document}

\title{Universal Approach to Quantum Adiabaticity via Ancilla Cavity}
\author{Lin Tian}
\email{ltian@ucmerced.edu}
\affiliation{School of Natural Sciences, University of California, Merced, California 95343, USA}

\begin{abstract}
A central challenge in the successful implementation of adiabatic quantum algorithms is to maintain the quantum adiabaticity during the entire evolution. However, the energy gap between the ground and the excited states of interacting many-body systems often decreases quickly with the number of qubits, and the quantum adiabaticity can be severely impaired. Despite numerous previous efforts, a practical method to preserve the quantum adiabaticity has yet to be developed. 
Here we present a universal approach to enhancing the quantum adiabaticity via cavity or circuit QED. By coupling an adiabatic quantum computer to an ancilla cavity, the coupled system can exhibit a bistable regime with bifurcation points, where the time evolution becomes very slow. Utilizing these generic nonlinear features, we show that the energy gap of the adiabatic quantum computer can be positioned between the bifurcation points, which results in strongly-enhanced quantum adiabaticity. We then apply this method to a quantum two-level system, an Exact Cover problem, and a transverse field Ising model. In contrast to previous works, this approach does not require the spectral knowledge of the quantum system or the construction of unphysical interactions and can be applied to a vast variety of adiabatic quantum processes.
\end{abstract}
\maketitle

\section{Introduction}
\label{sec:intro}

Adiabatic quantum computing is a powerful approach to generating desired many-body states and has been intensively studied for solving computationally-hard problems, such as the NP-complete problems and adiabatic optimization~\cite{FarhiScience2001, KadowakiPRE1998, Santoro2002, Boixo2014, Albash2016}. In a typical adiabatic quantum algorithm, the system Hamiltonian $H_{s}$ is tuned slowly from an initial Hamiltonian $H_{0}$ to a target Hamiltonian $H_{T}$ with $H_{s}(t)=[1-z(t)]H_{0}+z(t)H_{T}$, where $z(t)$ is a smooth function of the time $t$ with $z(t)=0$ at $t=0$ and $z(t)=1$ at the final time $T$~\cite{Farhi2000_1}. The system is initially prepared in the ground state of $H_{0}$, and the ground state of $H_{T}$ encodes the solution to the classically-intractable problem. During the time evolution, the system remains in the instantaneous ground state of $H_{s}(t)$ and evolves adiabatically to reach the ground state of $H_{T}$. 

To implement an adiabatic quantum algorithm successfully, the adiabatic criterion needs to be satisfied during the evolution. A commonly-adopted form of the adiabatic criterion is $\vert dH_{s}(t)/dt\vert \ll\Delta_{gp}^{2}$, where $\vert dH_{s}(t)/dt\vert$ is the magnitude of the transition matrix elements of $dH_{s}(t)/dt$ and $\Delta_{gp}$ is the energy gap between the ground and the excited states~\cite{AminPRL2009}. 
In an interacting many-body system, the energy gap often decreases quickly with the number of qubits, and the adiabatic criterion can be violated~\cite{MurgCirac2004}. The breaking of the quantum adiabaticity can induce diabatic transitions to the excited states with the system not reaching the desired ground state. For a quantum two-level system (TLS), the probability of such transitions is given by the Landau-Zener (LZ) formula~\cite{lz1,lz2}. In dynamical quantum phase transitions, such transitions have been widely studied in terms of the Kibble-Zurek mechanism and the scaling laws~\cite{DziarmagaAdvPhys2010, PolkovnikovRMP2011, ZurekPRL2005, PolkovnikovPRB2005, DziarmagaPRL2005}.

A number of methods have been developed to preserve the quantum adiabaticity and reduce the transition to the excited states. 
Some approaches aim at maintaining a finite energy gap via inhomogeneous ramping or by dividing the evolution into short segment~\cite{SchallerPRA2008, Farhi2002_2, ZurekDornerPTRSA2008, DziarmagaRamsNJP2010_1, delCampoPRL2010, ColluraPRL2010}. It was also shown that the energy gap can be enlarged by engineering the initial and the final Hamiltonian or by adding an intermediate Hamiltonian~\cite{DicksonPRL2011, Perdomo-OrtizQIP2011, FarhiQIP2011, VChoiPNAS2011, LZengJPA2016}. Spectral gap amplification has been designed in adiabatic quantum algorithms with frustration-free Hamiltonian~\cite{SommaSIAM2013}. In another method, the second-order phase transition was exploited to avoid exponentially-decreasing energy gap~\cite{SchuetzholdPRA}.
Other methods improve the local adiabaticity in the gap region with time-dependent ramping, including the nonlinear power-law ramping, the local adiabatic approach, and the optimal control approach~\cite{SenPRL2008, MondalPRB2009, BarankovPRL2008, RolandCerfPRA2002, HTQuanNJP2010, DoriaPRL2011, RahmaniPRL2011, PowerPRB2013, NWuPRB2015}. 
Meanwhile, geometric approaches that seek time-optimal path to the desired state have been investigated~\cite{RezakhaniPRL2009, ZulkowskiPRE2015}. Recently, it was shown that the desired ground state can be reached by adding counter-diabatic interactions to eliminate the transition to the excited states~\cite{XChenPRL2010, delCampoPRL2012, DeffnerPRX2014, Damski2014, MuthukrishnanPRX2016}.
However, all these approaches require either a priori knowledge of the energy spectrum of the quantum system, which is hard to obtain with classical methods, or the construction of unphysical interactions, which are challenging to implement in practical systems. 

Here we present a universal approach that can strongly enhance the quantum adiabaticity via cavity or circuit quantum electrodynamics (QED), where an adiabatic quantum computer is coupled to an ancilla cavity with $H_{0}$ as the coupling operator. The cavity serves as a knob that controls the Hamiltonian of the adiabatic quantum computer, but is also influenced by the state of the quantum computer. 
With the intrinsic nonlinear features of the operator average $X_{ss}=\langle H_{0}\rangle$ and its derivative $X_{ss}^{\prime}$, this coupled system can exhibit a bistable regime with bifurcation points, where the time evolution is very slow~\cite{DrazinBook1992, TianPRL2010, XWLuoPRA2016, TianPRA2016}. We show that the energy gap can be positioned between the bifurcation points so that the time evolution in the gap region can be significantly slowed down. We then conduct numerical simulation of this approach on a quantum TLS, an Exact Cover (EC) problem, and a one-dimensional transverse field Ising model (TFIM). Our numerical result shows that the quantum adiabaticity can be strongly enhanced. 

This approach is rooted in the generic properties of the operator average $X_{ss}$ and its derivative $X_{ss}^{\prime}$ in adiabatic quantum computers. We find that $X_{ss}$ increases monotonically but nonlinearly with the magnitude of $H_{0}$ in the Hamiltonian $H_{s}$, and furthermore, $X_{ss}^{\prime}$ reaches maximum in the gap region. These properties ensure the existence of a bistable regime with bifurcation points in this coupled system and make it possible to position the energy gap between the bifurcation points. 
More importantly, it only requires the knowledge of $X_{ss}$ and $X_{ss}^{\prime}$ at the initial and the target Hamiltonians to choose appropriate system parameters. In contrast to previous works, this method does not require the spectral knowledge of the adiabatic quantum computer, nor does it require the engineering of unphysical interactions. This approach can hence be applied to a vast variety of adiabatic quantum processes when combined with the cavity or circuit QED technology~\cite{circuitQED1, circuitQED2, DevoretScience2013, cavityQED, LangenAnnuRevCMP2015}. 

This paper is organized as follows. We first present our approach in Sec.~\ref{sec:ancillaapproach}. The generic properties of the operator average $X_{ss}$ and its derivation $X_{ss}\prime$, the stationary state and bifurcation points, and the requirements on the system parameters will be discussed in this section. We then apply this approach to a quantum TLS, an EC problem, and a TFIM in Secs.~\ref{sec:cavityTLS}, \ref{sec:cavityEC}, and \ref{sec:cavityTFIM}, respectively, and give the numerical result of the quantum adiabaticity for these models. In Sec.~\ref{sec:Discussions}, we discuss the generality of this approach with different control parameters and different forms of coupling. Conclusions are given in Sec.~\ref{sec:conclusions}.

\section{Adiabatic quantum computing with ancilla cavity}
\label{sec:ancillaapproach}

Consider an adiabatic quantum computer coupled to an ancilla cavity. The Hamiltonian of the adiabatic quantum computer is given by
\begin{equation}
H_{s}(B_{x})=-B_{x}H_{0}-J_{0}H_{T},\label{eq:Hs}
\end{equation}
where $B_{x}$ ($J_{0}$) is the magnitude of the Hamiltonian $H_{0}$ ($H_{T}$) with $B_{x}\gg J_{0}$. Different from previous works~\cite{FarhiScience2001, Farhi2000_1}, here $B_{x}$ and $J_{0}$ are fixed. The Hamiltonian of the cavity in the rotating frame of the driving field is
\begin{equation}
H_{c}=-\Delta_{c}a^{\dagger}a-\epsilon(a+a^{\dagger})+H_{cb},\label{eq:Hc}
\end{equation}
where $a$ ($a^{\dag}$) is the annihilation (creation) operator of the cavity mode, $\Delta_{c}$ is the detuning, $\epsilon$ is the driving amplitude, and $H_{cb}$ is the interaction between the cavity and its bath modes~\cite{TianPRL2007}. We assume the bath modes induce dissipation of the cavity with a damping rate $\kappa$. Let the coupling between the adiabatic quantum computer and the cavity be $H_{int}=g(a+a^{\dagger})H_{0}$ with a coupling strength $g$. This coupling is a dipole interaction between the cavity displacement $(a+a^{\dagger})$ and the Hamiltonian $H_{0}$~\cite{NoteCoupling, JXuePRB2017}, and has been widely studied in cavity and circuit QED~\cite{circuitQED1, circuitQED2, cavityQED}. The total Hamiltonian of this system is then $H_{t}=H_{s}+H_{c}+H_{int}$. 

We write $(a+a^{\dagger})=x_{a}+(\delta a+\delta a^{\dagger})$ in terms of the average cavity displacement $x_{a}=\langle a+a^{\dagger}\rangle$ and the fluctuation $\delta a$ ($\delta a^{\dagger}$). Similarly, $H_{0}=X+\delta H_{0}$ in terms of the operator average $X=\langle H_{0}\rangle$ and the fluctuation $\delta H_{0}$. Under strong driving and dissipation, the cavity can be treated semi-classically and the mean-field approximation can be applied~\cite{meanfield}. Under this approximation, the product of the fluctuations $g(\delta a+\delta a^{\dagger})\delta H_{0}$ in the interaction can be neglected with
\begin{equation}
H_{int}\approx gx_{a}H_{0}+gX(a+a^{\dagger})-gXx_{a}.\label{eq:HintMeanfield}
\end{equation}
The total Hamiltonian can then be decomposed as
\begin{equation}
H_{t} \approx H_{s}(\widetilde{B}_{x})+H_{c}(\widetilde{\epsilon}),\label{eq:HtApprox}
\end{equation}
where $H_{s}(\widetilde{B}_{x})$ is the Hamiltonian of the adiabatic quantum computer with an effective field $\widetilde{B}_{x}=B_{x}-gx_{a}$, and $H_{c}(\widetilde{\epsilon})$ is the Hamiltonian of the ancilla cavity with an effective driving amplitude $\widetilde{\epsilon}=\epsilon-gX$. The constant term $-gXx_{a}$ in (\ref{eq:HintMeanfield}) is omitted from (\ref{eq:HtApprox}). 
Here the Hamiltonian $H_{s}(\widetilde{B}_{x})$ depends on the average cavity displacement $x_{a}$ and the Hamiltonian $H_{c}(\widetilde{\epsilon})$ depends on the operator average $X=\langle H_{0}\rangle$. 
The dynamics of the adiabatic quantum computer is governed by the Schr\"{o}dinger equation: 
\begin{equation}
id\left|\psi_{s}\right\rangle /dt=H_{s}(\widetilde{B}_{x})\left|\psi_{s}\right\rangle \label{eq:SchroedingerEq}
\end{equation}
with $|\psi_{s}\rangle $ being the state of the quantum computer. And the dynamics of the cavity is governed by the Heisenberg-Langevin equation 
\begin{equation}
d\left\langle a\right\rangle /dt  =  i\Delta_{c}\left\langle a\right\rangle -\frac{\kappa}{2}\left\langle a\right\rangle +i\widetilde{\epsilon}.\label{eq:dadt}
\end{equation}
Both stationary and dynamical behaviors of this coupled system can be obtained by solving these equations self-consistently.

\subsection{Generic properties of $X_{ss}$ and $X_{ss}^{\prime}$}
\label{sub:Xproperty}

We define $X_{ss}(\widetilde{B}_{x})=\langle\psi_{G}\vert H_{0}\vert\psi_{G}\rangle$ and denote $X_{ss}^{\prime}(\widetilde{B}_{x})$ as its derivative, where $\vert\psi_{G}\rangle$ is the ground state of the Hamiltonian $H_{s}(\widetilde{B}_{x})$. Our approach to enhancing the quantum adiabaticity strongly relies on the generic nonlinear features of $X_{ss}$ and $X_{ss}^{\prime}$ in adiabatic quantum computers. 
Below we calculate the derivative $X_{ss}^{\prime}$ using the relation $X_{ss}^{\prime}(\widetilde{B}_{x})=[X_{ss}(\widetilde{B}_{x}+\delta B_{x})-X_{ss}(\widetilde{B}_{x})]/\delta B_{x}$ with $X_{ss}(\widetilde{B}_{x}+\delta B_{x})=\langle\psi_{G}^{\delta}\vert H_{0}\vert\psi_{G}^{\delta}\rangle$ and $\delta B_{x}\rightarrow0$. Here $\vert\psi_{G}^{\delta}\rangle$ is the ground state of the Hamiltonian $H_{s}(\widetilde{B}_{x}+\delta B_{x})$ and can be written as $\vert\psi_{G}^{\delta}\rangle=\sum c_{n}\vert\psi_{n}\rangle$ with $\vert\psi_{n}\rangle$ being an eigenstate of $H_{s}(\widetilde{B}_{x})$ and $c_{n}$ being an overlap coefficient. We find that $X_{ss}^{\prime}(\widetilde{B}_{x})=\sum_{n\ne G}(\vert c_{n}\vert/\delta B_{x})^{2}[E_{n}-E_{G}]$ with $E_{n}$ being the eigenenergy of $\vert\psi_{n}\rangle$. Assuming a finite energy gap with $(E_{n}-E_{G})>0$ and using the second order perturbation theory, we obtain 
\begin{equation}
X_{ss}^{\prime}(\widetilde{B}_{x})=\sum_{n\ne G}\frac{\left|\left\langle \psi_{n}\vert H_{0}\vert\psi_{G}\right\rangle \right|^{2}}{E_{n}-E_{G}}.\label{eq:dXdBxSimp}
\end{equation}
Therefore, $X_{ss}^{\prime}>0$ in all parameter regimes and  $X_{ss}$ increases monotonically with $\widetilde{B}_{x}$. 

In adiabatic quantum algorithms, the eigenbasis of the Hamiltonian $H_{T}$ ($H_{0}$) is typically made of eigenstates of the Pauli operators $\sigma_{zi}$ ($\sigma_{xi}$) of the qubits. At $\widetilde{B}_{x}=0$, $H_{s}=-J_{0}H_{T}$, where the energy separation $E_{n}-E_{G}=O(J_{0})$ and the matrix element $\langle \psi_{n}|H_{0}|\psi_{G}\rangle \sim O(1)$ for the low-lying excited states. This analysis shows that at $\widetilde{B}_{x}=0$, $X_{ss}=0$ and $X_{ss}^{\prime}=O(N)/J_{0}$. At $\widetilde{B}_{x}=B_{x}$ with $B_{x}\gg J_{0}$, $H_{s}\approx-\widetilde{B}_{x}H_{0}$, where $E_{n}-E_{G}=O(B_{x})$ and $\langle \psi_{n}\vert H_{0}\vert\psi_{G}\rangle \rightarrow0$ for the low-lying excited states. We then have $X_{ss}=O(N)$ and $X_{ss}^{\prime}\rightarrow0$ at $\widetilde{B}_{x}=B_{x}$. When $\widetilde{B}_{x}$ is near the position of the energy gap, $\langle \psi_{n} |H_{0}|\psi_{G}\rangle \sim O(1)$, but the energy separation $E_{n}-E_{G}\sim \Delta_{gp}$ for the low-lying excited states can be much smaller than $J_{0}$. From (\ref{eq:dXdBxSimp}), we can deduce that $X_{ss}^{\prime}$ reaches maximum in the gap region. The above properties of $X_{ss}$ and $X_{ss}^{\prime}$ are generic in adiabatic quantum computers, as confirmed by our result in Fig.~\ref{fig3}(b) and Fig.~\ref{fig4}(a).

\subsection{Stationary state and bifurcation points}
\label{sub:Stationary-state}

\begin{figure}
\includegraphics[clip,width=8.5cm]{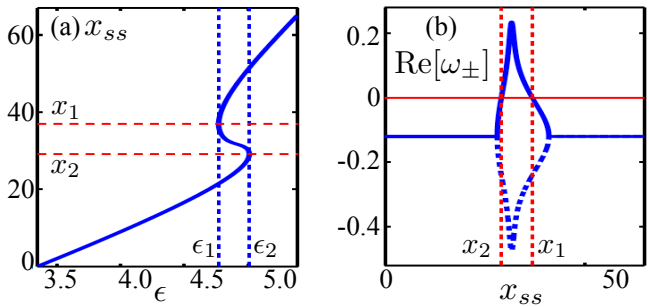}
\caption{(a) $x_{ss}$ vs $\epsilon$ for a TFIM of $N=120$ qubits.
(b) $\textrm{Re}[\omega_{+}]$ (solid) and $\textrm{Re}[\omega_{-}]$ (dashed) vs $x_{ss}$. 
Thin dashed lines indicate the driving amplitudes $\epsilon_{1,2}$ and the displacements $x_{1,2}$ at the bifurcation points. 
The parameters are $J_{0}=1$, $B_{x}=1.95$, $\Delta_{c}=-0.14$, $\kappa=0.12$, and $g=0.03$.}
\label{fig1} 
\end{figure}
For given parameters with the adiabatic quantum computer in its ground state, this system can reach a stationary state with $d\langle a\rangle /dt=0$. From (\ref{eq:dadt}), we derive the stationary cavity displacement as $x_{ss}=(\epsilon-gX_{ss}(\widetilde{B}_{x}))/\alpha$ with $\alpha=-[\Delta_{c}^{2}+(\kappa/2)^{2}]/2\Delta_{c}$ and $\widetilde{B}_{x}=B_{x}-gx_{ss}$. Here $x_{ss}$ and $X_{ss}(\widetilde{B}_{x})$ depend mutually on each other and can be solved self-consistently. For illustration, we plot the displacement $x_{ss}$ of a cavity coupled to a TFIM~\cite{SachdevBook1999} vs the driving amplitude $\epsilon$ in Fig.~\ref{fig1}(a). Details of this model are given in Sec.~\ref{sec:cavityTFIM}.
The solution exhibits a bistable regime, where two stable and one unstable solutions exist at a given driving amplitude with the bifurcation points $\epsilon_{1}$ and $\epsilon_{2}$. This is a universal feature of an adiabatic quantum computer coupled to a cavity resulting from the nonlinear properties of $X_{ss}$ and $X_{ss}^{\prime}$. 

It can be shown that $d\epsilon/dx_{ss}=\alpha-g^{2}X_{ss}^{\prime}(\widetilde{B}_{x})$. At the bifurcation points, $d\epsilon/dx_{ss}=0$, i.e., $X_{ss}^{\prime}(\widetilde{B}_{x})=\alpha/g^{2}$, as indicated by the thin dashed lines in Fig.~\ref{fig1}(a). Hence, for this system to contain two bifurcation points, the parameters need to satisfy the condition $X_{ss}^{\prime}(0)<\alpha/g^{2}<\max[X_{ss}^{\prime}]$. 

The dynamics near the bifurcation points is strongly affected by the above nonlinear features. Define a small shift $\delta x_{a}$ ($\delta p_{a}$) from the stationary displacement (momentum) with $x_{a}=x_{ss}+\delta x_{a}$ ($p_{a}=p_{ss}+\delta p_{a}$). Assume the adiabatic quantum computer remains in the ground state so that the operator average can be linearized as $X=X_{ss}-gX_{ss}^{\prime}\delta x_{a}$. Using (\ref{eq:dadt}), we obtain 
\begin{equation}
\frac{d}{dt}\left[\begin{array}{c}
\delta x_{a}\\
\delta p_{a}
\end{array}\right]=\left[\begin{array}{cc}
-\kappa/2 & -\Delta_{c}\\
\Delta_{c}-2gX_{ss}^{\prime} & -\kappa/2
\end{array}\right]\left[\begin{array}{c}
\delta x_{a}\\
\delta p_{a}
\end{array}\right].\label{eq:ddx_ddp}
\end{equation}
The secular frequencies of the small shift are then
\begin{equation}
\omega_{\pm}=-\kappa/2\pm\sqrt{\left(-2\Delta_{c}g^{2}X_{ss}^{\prime}-\Delta_{c}^{2}\right)}.\label{eq:om12}
\end{equation}
At the bifurcation points with $X_{ss}^{\prime}(\widetilde{B}_{x})=\alpha/g^{2}$, the real part of $\omega_{+}$ approaches zero, as shown in Fig.~\ref{fig1}(b). A vanishing real part indicates that the time evolution in the vicinity of the bifurcation points becomes very slow. During an adiabatic evolution, the quantum adiabaticity in the gap region is the most vulnerable to diabatic transitions. In our approach, using the knowledge of $X_{ss}$ and $X_{ss}^{\prime}$ at $\widetilde{B}_{x}=0$ and $B_{x}$, we can engineer the system parameters to position the energy gap between the bifurcation points so as to slow down the time evolution in this region and enhance the quantum adiabaticity.

\subsection{Adiabatic protocol}
\label{sub:Parameters-and-protocol}

Below we choose the driving amplitude as the control parameter in our adiabatic protocol. From the discussions in Secs.~\ref{sub:Xproperty} and \ref{sub:Stationary-state}, the parameters of this system need to satisfy the following conditions. First, $\Delta_{c}<0$, i.e., $\alpha>0$, so that the real part of the secular frequency $\omega_{+}$ can reach zero. Second, $X_{ss}^{\prime}(0)<\alpha/g^{2}<\max(X_{ss}^{\prime})$ so that there are at least two bifurcation points in the driving amplitude. Finally, $\epsilon_{i}<\epsilon_{f}$, i.e., $X_{ss}(B_{x})/B_{x}<\alpha/g^{2}$, so that the system switches from the lower branch to the upper branch (or vice versa) at a bifurcation point and crosses the gap region during the switching. In addition to these conditions, the cavity damping rate $\kappa$ and cavity detuning $\Delta_{c}$ strongly affect the dynamics of this system. They can be chosen based on the magnitudes of $B_{x}$ and $J_{0}$ and on the realistic range of the parameters in a physical system. 
To determine the system parameters, it only requires the knowledge of $X_{ss}$ and $X_{ss}^{\prime}$ at $\widetilde{B}_{x}=0$ and $B_{x}$, which can be obtained analytically. Our approach thus does not require the spectral knowledge of the adiabatic quantum computer or the engineering of unphysical interactions. 

As a control parameter, the driving amplitude can be tuned slowly or suddenly. In both cases, the system will evolve continuously~\cite{TianPRA2016}. We adopt the following adiabatic protocol for our numerical simulation in Secs.~\ref{sec:cavityTLS}, \ref{sec:cavityEC}, and \ref{sec:cavityTFIM}. The driving amplitude is initially biased at $\epsilon_{0}$ with $x_{ss}=0$ ($\widetilde{B}_{x}=B_{x}$) and the adiabatic quantum computer is in the ground state of $H_{s}(B_{x})$. At time $t=0$, the driving amplitude is switched to an intermediate value $\epsilon$ and the system starts evolving towards the stationary state of $\epsilon$. For $\epsilon$ larger than a value $\epsilon_{c}\approx\epsilon_{2}$, the effective field $\widetilde{B}_{x}$ decreases to cross the position of the energy gap. When $\widetilde{B}_{x}$ is well below the gap position, the driving amplitude is tuned to $\epsilon_{f}$, which corresponds to $x_{ss}=B_{x}/g$ ($\widetilde{B}_{x}=0$). In the ideal scenario, the system then evolves towards the ground state of the Hamiltonian $-J_{0}H_{T}$. 

During the evolution, diabatic transitions to the excited states can occur, which will degrade the fidelity of the final state. The aim of our protocol is to find an $\epsilon$ that greatly reduces the probability of such transitions. Because we do not have accurate knowledge of $\epsilon_{c}$ ($\epsilon_{2}$), the above protocol needs to be repeated a number of times to reach a desired $\epsilon$. The effectiveness of this protocol can be characterized by the difference between the final field $\widetilde{B}_{x}(T)$ and the ideal value $\widetilde{B}_{x}=0$, which is caused by the transitions to the excited states. In the following, we illustrate this approach with three models.

\section{Quantum two-level system}
\label{sec:cavityTLS}

The Hamiltonian of the TLS is given by (\ref{eq:Hs}) with $H_{T} = -B_{x}\sigma_{x}/2J_{0}+\sigma_{z}/2$ and $H_{0}=\sigma_{x}$. Here $\sigma_{x,z}$ are the Pauli matrices of the TLS and $B_{x}\gg J_{0}$. Different from typical adiabatic quantum computers, $H_{0}$ and $H_{T}$ almost share the same set of eigenstates $\vert\pm\rangle=(\vert0\rangle\pm\vert 1\rangle)/\sqrt{2}$, where $\vert0\rangle$ and $\vert1\rangle$ are eigenstates of $\sigma_{z}$. The cavity Hamiltonian is (\ref{eq:Hc}) and the coupling between the TLS and the cavity is $H_{int}=g(a+ a^{\dagger}) H_{0}$. As discussed in Sec.~\ref{sec:ancillaapproach}, the total Hamiltonian of this system can be decomposed into the TLS Hamiltonian $H_{s}(\widetilde{B}_{x})$ with an effective field $\widetilde{B}_{x}=B_{x}-gx_{a}$ and the cavity Hamiltonian $H_{c}(\widetilde{\epsilon})$ with an effective
driving $\widetilde{\epsilon}=\epsilon-gX$ under the mean-field approximation. When $\widetilde{B}_{x}$ is swept from $B_{x}$ to $0$, the magnitude of the $\sigma_{x}$-component in $H_{s}(\widetilde{B}_{x})$ is swept from $B_{x}/2$ to $-B_{x}/2$ with the magnitude of the $\sigma_{z}$-component unchanged. The energy gap occurs at $\widetilde{B}_{x}=B_{x}/2$, where $H_{s}(\widetilde{B}_{x})=-J_{0}\sigma_{z}/2$ and $\Delta_{gp}=J_{0}$.
\begin{figure}
\includegraphics[clip,width=8.5cm]{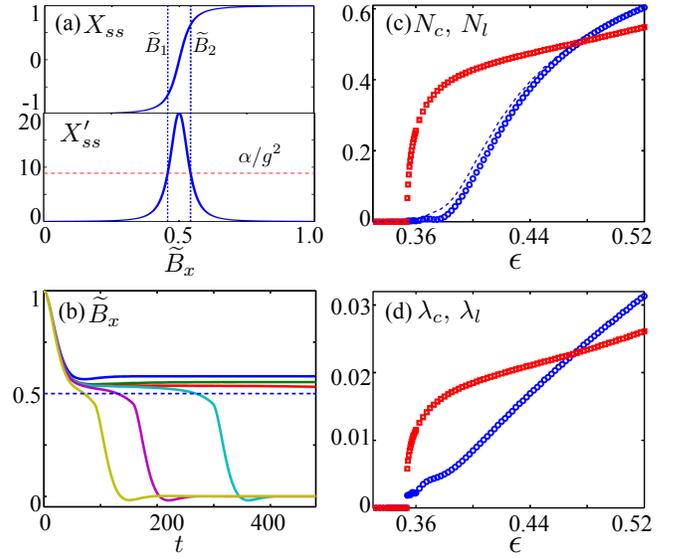}
\caption{(a) $X_{ss}$ and $X_{ss}^{\prime}$ vs $\widetilde{B}_{x}$. Dashed lines corresponds to $X_{ss}^{\prime}=\alpha/g^{2}$ and $\widetilde{B}_{x}=\widetilde{B}_{1,2}$ at the bifurcation points. 
(b) $\widetilde{B}_{x}$ vs the time $t$ at $\epsilon=0.34,\,0.35,\,0.3535,\,0.354,\,0.36,\,0.37$ from top to bottom. The dashed line corresponds to $\widetilde{B}_{x}$ at the gap position. 
(c) $N_{c}$ from the numerical simulation (circles) and the Landau-Zener formula (dashed line) and $N_{l}$ in the linear-ramping model (squares) vs $\epsilon$. 
(d) $\lambda_{c}$ of the cavity-coupled TLS (circles) and $\lambda_{l}$ of the linear-ramping model (squares) vs $\epsilon$.
Other parameters are $J_{0}=0.1$ , $B_{x}=1$, $\kappa=0.1$, $\Delta_{c}=-0.05$, and $g=0.075$.}
\label{fig2}
\end{figure}

The ground-state average of the coupling operator $H_{0}$ can be derived as
\begin{equation}
X_{ss}(\widetilde{B}_{x})=(2\widetilde{B}_{x}-B_{x})/\sqrt{(2\widetilde{B}_{x}-B_{x})^{2}+J_{0}^{2}}.\label{eq:XssTLS}
\end{equation}
At $\widetilde{B}_{x}=B_{x}$ with $B_{x}\gg J_{0}$, $X_{ss}\approx1$ and $X_{ss}^{\prime}\rightarrow0$. At $\widetilde{B}_{x}=0$, $X_{ss}\approx-1$ and $X_{ss}^{\prime}\rightarrow0$. The maximum of the derivative occurs at the gap position with $\max[X_{ss}^{\prime}]=2/J_{0}$. Both $X_{ss}$ and $X_{ss}^{\prime}$ are plotted in Fig.~\ref{fig2}(a). 
With the requirements on the parameters discussed in Sec.~\ref{sub:Parameters-and-protocol}, we choose $J_{0}=0.1$, $B_{x}=1$, $\kappa=0.1$, $\Delta_{c}=-0.05$, and $g=0.075$. These parameters yield $\alpha/g^{2}=8.9$ with two bifurcation points at $\epsilon_{1}=0.31$ and $\epsilon_{2}=0.35$. The corresponding effective fields at the bifurcation points are $\widetilde{B}_{1}=0.48$ and $\widetilde{B}_{2}=0.54$. Meanwhile, at $\epsilon_{0}=0.07$, $x_{ss}=0$ with $\widetilde{B}_{x}=B_{x}$; and at $\epsilon_{f}=0.59$, the displacement reaches maximum with $x_{ss}=B_{x}/g$ and $\widetilde{B}_{x}=0$. 

We simulate the time evolution of this system numerically following the protocol outlined in Sec.~\ref{sub:Parameters-and-protocol}. In Fig.~\ref{fig2}(b), the effective field $\widetilde{B}_{x}$ is plotted vs the time $t$ at several values of $\epsilon$. For $\epsilon$ below a value $\epsilon_{c}\approx \epsilon_{2}$, $\widetilde{B}_{x}$ remains above the gap position $B_{x}/2$ during the entire evolution. For $\epsilon$ above $\epsilon_{c}$, $\widetilde{B}_{x}$ decreases to cross the gap position. When $\widetilde{B}_{x}$ is below $(B_{x}-J_{0})/2$, where the energy separation between the ground and the excited states is sufficiently large to prevent further diabatic effects, the driving amplitude is switched to $\epsilon_{f}$. When $\epsilon$ approaches $\epsilon_{c}$, the time evolution of the effective field in the gap region slows down significantly. Similar slowdown can be observed when the driving amplitude is switched down from $\epsilon_{f}$ to $\epsilon_{0}$. This result indicates that the quantum adiabaticity of this process can be enhanced by choosing appropriate driving amplitude $\epsilon$. 

The quantum adiabaticity in the above process can be characterized with the ramping rate at the gap position $\lambda_{c}=\vert d\widetilde{B}_{x}/dt\vert$~\cite{TianPRA2016}. For convenience of discussion, we let $\lambda_{c}=0$ if $\widetilde{B}_{x}(t)$ does not cross the gap position. 
The numerical result of $\lambda_{c}$ can be tested by estimating the probability of the TLS in the excited state with the Landau-Zener formula $N_{c} = \exp(-\pi\Delta_{qp}^{2}/2\lambda_{c})$~\cite{lz1,lz2}. In Fig.~\ref{fig2}(c), the estimated probability is compared with the probability obtained from the numerical simulation. The two results demonstrate excellent agreement, which verifies that $\lambda_{c}$ is a good index to study the quantum adiabaticity in this system. As shown in Fig.~\ref{fig2}(d), $\lambda_{c}$ decreases significantly as $\epsilon$ approaches $\epsilon_{c}$. For comparison, we define a linear-ramping model with a duration $t_{s}$ and a linear-ramping rate $\lambda_{l}=\vert\widetilde{B}_{x}(T)-\widetilde{B}_{x}(0)\vert/t_{s}$, where $t_{s}$ is the time in the above numerical simulation for the effective field to reach $\widetilde{B}_{x}(T)$~\cite{ZurekPRL2005, PolkovnikovPRB2005}. Our result reveals that in a wide range of the driving amplitude $\epsilon$, our approach outperforms the linear-ramping model with $\lambda_{c}<\lambda_{l}$ and demonstrates strongly-enhanced quantum adiabaticity. As a result, in Fig.~\ref{fig2}(c), the excitation probability is greatly reduced with $N_{c}<N_{l}$.

\section{Exact cover problem}
\label{sec:cavityEC}

The EC problem is NP-complete and has been intensively studied in adiabatic quantum computing~\cite{FarhiScience2001, Albash2016}. Here we randomly generate an EC instance on $N=6$ qubits. This instance contains $m=5$ clauses:
\begin{eqnarray}
C_{1} & = & (Q_{1},\, Q_{2},\, Q_{5}),\nonumber \\
C_{2} & = & (Q_{2},\, Q_{3},\, Q_{6}),\nonumber \\
C_{3} & = & (Q_{3},\, Q_{4},\, Q_{6}),\label{eq:clauses}\\
C_{4} & = & (Q_{1},\, Q_{3},\, Q_{5}),\nonumber \\
C_{5} & = & (Q_{2},\, Q_{5},\, Q_{6}),\nonumber 
\end{eqnarray}
where $Q_{j}=0$ ($1$) corresponds to the eigenstate $\vert0\rangle$ ($\vert1\rangle$) of the Pauli operator $\sigma_{zj}$ for the $j$th qubit. The clause $C_{i}$ depends on the states of three qubits $(Q_{i1},\, Q_{i2},\, Q_{i3})$ and is satisfied when one and only one of these states is $1$ with the other two in the state $0$, i.e., $Q_{i1}+Q_{i2}+Q_{i3}=1$. The instance in (\ref{eq:clauses}) has a unique solution: $Q_{1}=Q_{6}=1$ and all other states are $0$.
The Hamiltonian that encodes the solution to this instance can be written as $H_{T}=\sum_{i}H_{Ti}$ with $H_{Ti}=\sum f_{i}(\vec{Q})\vert\vec{Q}\rangle\langle\vec{Q}\vert$ and $\vert\vec{Q}\rangle=\prod_{j}\vert Q_{j}\rangle$ being a product state of all qubits. Here $f_{i}(\vec{Q}) =0$ if the clause $C_{i}$ is satisfied and $f_{i}(\vec{Q})=1$ if $C_{i}$ is violated by the state $\vec{Q}$. From this definition, $H_{T}=\sum f(\vec{Q})\vert\vec{Q}\rangle\langle\vec{Q}\vert$ with $f(\vec{Q})$ being the number of violated clauses by the state $\vec{Q}$. The ground state of this Hamiltonian satisfies all clauses and has $f(\vec{Q})=0$.
\begin{figure}
\includegraphics[clip,width=8.5cm]{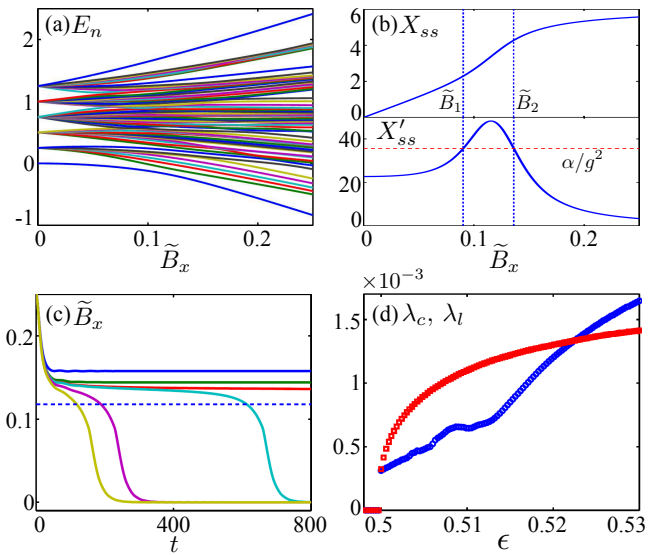}
\caption{(a) The energy spectrum of the EC instance (\ref{eq:clauses}) vs $\widetilde{B}_{x}$. 
(b) $X_{ss}$ and $X_{ss}^{\prime}$ vs $\widetilde{B}_{x}$. Dashed lines corresponds to $X_{ss}^{\prime}=\alpha/g^{2}$ and $\widetilde{B}_{x}=\widetilde{B}_{1,2}$ at the bifurcation points. 
(c) $\widetilde{B}_{x}$ vs the time $t$ at $\epsilon=0.49,\,0.498,\,0.499,\,0.5,\,0.502,\,0.51$ from top to bottom. The dashed line corresponds to $\widetilde{B}_{x}$ at the gap position. 
(d) $\lambda_{c}$ of the cavity-coupled TLS (circles) and $\lambda_{l}$ of the linear-ramping model (squares) vs $\epsilon$.
Other parameters are $J_{0}=0.25$ , $B_{x}=0.5$, $\kappa=0.25$, $\Delta_{c}=-0.1$, and $g=0.06$.}
\label{fig3} 
\end{figure}

In our adiabatic protocol, the Hamiltonian of the above EC instance is given by (\ref{eq:Hs}), where $H_{0}=\sum_{j}\sigma_{xj}$ with $\sigma_{xj}$ being a Pauli operator of the $j$th qubit and $J_{0}<0$. With $B_{x}\gg \vert J_{0}\vert$, $H_{s}(B_{x})\approx-B_{x}H_{0}$, and the ground state can be approximated as $\Pi_{j}\vert+\rangle$ with all qubits in the state $\vert+\rangle=(\vert0\rangle+\vert1\rangle)/ \sqrt{2}$. The cavity Hamiltonian is given by (\ref{eq:Hc}) and the qubits are coupled to the cavity via the interaction $H_{int}=g(a+a^{\dagger})H_{0}$. The total Hamiltonian can be decomposed into the EC part $H_{s}(\widetilde{B}_{x})$ with $\widetilde{B}_{x}=B_{x}-gx_{a}$ and the cavity part $H_{c}(\widetilde{\epsilon})$ with $\widetilde{\epsilon}=\epsilon-gX$ under the mean-field approximation, as discussed in Sec.~\ref{sec:ancillaapproach}. Here the effective field $\widetilde{B}_{x}$ can be tuned continuously from $B_{x}$ to $0$ by varying the cavity state. The energy spectrum of $H_{s}(\widetilde{B}_{x})$ is plotted in Fig.~\ref{fig3}(a) for $J_{0}=0.25$ and $B_{x}=0.5$. From the numerical result, we find that the energy gap is at $\widetilde{B}_{x}=0.12$ with $\Delta_{gp}=0.10$. 

It can be shown that the ground-state operator average $X_{ss}=0$ at $\widetilde{B}_{x}=0$ and $X_{ss}\rightarrow N$ at $\widetilde{B}_{x}\gg J_{0}$. With a perturbation theory approach, we obtain that $X_{ss}^{\prime}=22.7$ at $B_{x}=0$ and $X_{ss}^{\prime}\rightarrow0$ at $\widetilde{B}_{x}\gg J_{0}$. In Fig.~\ref{fig3}(b), $X_{ss}$ and $X_{ss}^{\prime}$ are plotted, which exhibit the universal features discussed in Sec.~\ref{sub:Xproperty}. We then choose the parameters for the cavity and the coupling with $\kappa=0.25$, $\Delta_{c}=-0.1$, and $g=0.06$. These parameters yield $\alpha/g^{2}=35.6$ with two bifurcation points at $\epsilon_{1}=0.48$ and $\epsilon_{2}=0.5$, respectively. Meanwhile, at $\epsilon_{0}=0.33$, $x_{ss}=0$; and at $\epsilon_{f}=0.53$, the cavity displacement reaches its maximum with $x_{ss}=B_{x}/g$. 

Following the switching protocol described in Sec.~\ref{sub:Parameters-and-protocol}, we simulate the dynamics of this coupled system by solving (\ref{eq:SchroedingerEq}) and (\ref{eq:dadt}). Here the driving amplitude is switched to $\epsilon_{f}$ when the effective field is below $0.09$. The time dependence of $\widetilde{B}_{x}$ is given in Fig.~\ref{fig3}(c), which demonstrates similar behaviors as those in Fig.~\ref{fig2}(b). In particular, the time variation of $\widetilde{B}_{x}$ in the gap region significantly slows down when the driving amplitude $\epsilon$ approaches a value $\epsilon_{c}\approx\epsilon_{2}$. 
In Fig.~\ref{fig3}(d), the ramping rate $\lambda_{c}$ is plotted vs $\epsilon$ together with the linear-ramping rate $\lambda_{l}$ defined in Sec.~\ref{sec:cavityTLS}. Both $\lambda_{c}$ and $\lambda_{l}$ decrease quickly when the driving amplitude $\epsilon$ approaches $\epsilon_{c}$ and $\lambda_{c}<\lambda_{l}$ in a wide range of $\epsilon$. This result demonstrates strong enhancement of the quantum adiabaticity using our approach in comparison with the linear-ramping model.

\section{Transverse field Ising model}
\label{sec:cavityTFIM}

The Hamiltonian of a one-dimensional TFIM can be written as (\ref{eq:Hs}) with $H_{0}=\sum_{i}\sigma_{xi}$ and $H_{T}=\sum_{i}\sigma_{zi}\sigma_{zi+1}$~\cite{SachdevBook1999}. Here $B_{x}$ is the transverse magnetic field applied to the qubits, $J_{0}$ is the ferromagnetic coupling between adjacent qubits, $\sigma_{xi}$, $\sigma_{zi}$ are the Pauli operators of the $i$th qubit, and $N$ is the total number of qubits in the chain. The Hamiltonian of the cavity is given by (\ref{eq:Hc}) and the TFIM is coupled to the cavity via the dipole interaction $H_{int}=g(a+a^{\dagger})H_{0}$. Under the mean-field approximation, the total Hamiltonian can be decomposed into the TFIM part $H_{s}(\widetilde{B}_{x})$ with $\widetilde{B}_{x} =B_{x}-gx_{a}$ and the cavity part $H_{c}(\widetilde{\epsilon})$ with $\widetilde{\epsilon}=\epsilon-gX$. The energy spectrum of $H_{s}(\widetilde{B}_{x})$ is exactly solvable using the Jordan-Wigner transformation~\cite{JWT}. At $\widetilde{B}_{x}\gg J_{0}$ ($\widetilde{B}_{x}\ll J_{0}$), the ground state of the TFIM is in a paramagnetic (ferromagnetic) phase. The energy gap occurs at the critical point $\widetilde{B}_{x}=J_{0}$ with $\Delta_{gp}\approx2\pi J_{0}/N$ for large $N$. 

\begin{figure}
\includegraphics[clip,width=8.5cm]{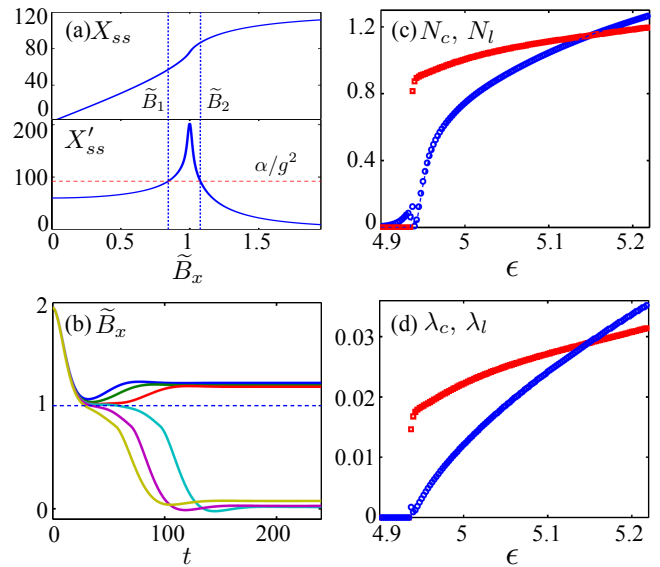}
\caption{(a) $X_{ss}$ and $X_{ss}^{\prime}$ vs $\widetilde{B}_{x}$. Dashed lines corresponds to $X_{ss}^{\prime}=\alpha/g^{2}$ and $\widetilde{B}_{x}=\widetilde{B}_{1,2}$ at the bifurcation points.
(b) $\widetilde{B}_{x}$ vs the time $t$ at $\epsilon=4.90,\,4.92,\,4.935,\,4.938,\,4.95,\,4.97$ from top to bottom. The dashed line corresponds to $\widetilde{B}_{x}$ at the gap position. 
(c) $N_{c}$ from the numerical simulation (circles) and the Landau-Zener formula (dashed line) and $N_{l}$ in the linear-ramping model (squares) vs $\epsilon$. 
(d) $\lambda_{c}$ of the cavity-coupled TLS (circles) and $\lambda_{l}$ of the linear-ramping model (squares) vs $\epsilon$. 
Other parameters are the same as those in Fig.~\ref{fig1}.}
\label{fig4}
\end{figure}

In this model, the ground-state operator average $X_{ss}=0$ at $\widetilde{B}_{x}=0$ and $X_{ss}\rightarrow N$ at $\widetilde{B}_{x}\gg J_{0}$. Using a perturbation theory approach, we find that $X_{ss}^{\prime}=N/2J_{0}$ at $\widetilde{B}_{x}=0$ and $X_{ss}^{\prime}\rightarrow0$ at $\widetilde{B}_{x}\gg J_{0}$. The maximum of $X_{ss}^{\prime}$ appears at the critical point and diverges in the thermodynamic limit. Both $X_{ss}$ and $X_{ss}^{\prime}$ are shown in Fig.~\ref{fig4}(a). We choose the following parameters: $B_{x}=1.95$, $J_{0}=1$, $\kappa=0.12$, $\Delta_{c}=-0.14$, and $g=0.03$. These parameters yield $\alpha/g^{2}=92.1$ with two bifurcation points at $\epsilon_{1}=4.77$ and $\epsilon_{2}=5.01$, respectively. We also obtain $x_{ss}=0$ at $\epsilon_{0}=2.23$ and the cavity displacement reaches its maximum $x_{ss}=B_{x}/g$ at $\epsilon_{f}=5.38$. 

For an initial state in the subspace of states with even number of excitations at a given effective field, the wave function of the TFIM at a time $t$ can be written as $|\psi(t)\rangle =\prod(U_{k}(t)+iV_{k}(t)c_{k}^{\dagger}c_{-k}^{\dag})|0\rangle$ with time-dependent coefficients $U_{k}(t)$ and $V_{k}(t)$. Here $k=(2m-1)\pi/N$ with $1\le m\le N/2$, $c_{k}$ ($c_{k}^{\dag}$) is the annihilation (creation) operator of a fermionic particle at the quasimomentum $k$, and $|0\rangle$ is the vacuum state~\cite{DziarmagaPRL2005}. The time evolution of $\vert\psi(t)\rangle $ is governed by (\ref{eq:SchroedingerEq}), which can be converted to
\begin{equation}
i\left(\begin{array}{c}
\dot{\bar{U}}_{k}\\
\dot{\bar{V}}_{k}
\end{array}\right)=\left(\begin{array}{cc}
-\varepsilon_{k}\cos2\theta_{k} & -\varepsilon_{k}\sin2\theta_{k}\\
-\varepsilon_{k}\sin2\theta_{k} & +\varepsilon_{k}\cos2\theta_{k}
\end{array}\right)\left(\begin{array}{c}
\bar{U}_{k}\\
\bar{V}_{k}
\end{array}\right)\label{eq:dUV}
\end{equation}
in the subspace of the quasimomenta $\pm k$. Here $\bar{U}_{k}=U_{k}e^{i\theta_{k}}$, $\bar{V}_{k}=V_{k}e^{i\theta_{k}}$, $\theta_{k}$ is an overall phase factor, and $\varepsilon_{k}=2(J_{0}^{2}+\widetilde{B}_{x}^{2}-2\widetilde{B}_{x}J_{0}\cos k)^{1/2}$ is the quasiparticle energy for the time-dependent effective field $\widetilde{B}_{x}$. The dynamics of this system can be obtained by solving (\ref{eq:dadt}) and (\ref{eq:dUV}) self-consistently. 

Under the adiabatic protocol given in Sec.~\ref{sub:Parameters-and-protocol}, we simulate the time evolution of this system numerically. During the evolution, the driving amplitude is tuned to $\epsilon_{f}$ when the effective field decreases below $\widetilde{B}_{x}=0.8$. The time dependence of $\widetilde{B}_{x}$ is plotted in Fig.~\ref{fig4}(b) for several driving amplitudes, which exhibits similar behaviors to that in Fig.~\ref{fig2}(b) and Fig.~\ref{fig3}(c). When $\epsilon$ approaches a value $\epsilon_{c}\approx \epsilon_{2}$, the time variation of $\widetilde{B}_{x}$ in the gap region becomes very slow. 
The probability of excitation can be defined as $N_{c}=\sum_{k>0}|\beta_{k}|^{2}$, where $\beta_{k}$ is the probability amplitude of the excited state in the $\pm k$ subspace~\cite{NoteWavefunction}. In Fig.~\ref{fig4}(c), $N_{c}$ is plotted vs the driving amplitude $\epsilon$. As $\epsilon$ approaches $\epsilon_{c}$, $N_{c}$ decreases accordingly. The corresponding probability of excitation $N_{l}$ in the linear-ramping model defined in Sec.~\ref{sec:cavityTLS} also decreases, but $N_{c}<N_{l}$ in a wide range of $\epsilon$. In Fig.~\ref{fig4}(d), the ramping rate $\lambda_{c}$ is plotted together with the linear-ramping rate $\lambda_{l}$. With $\epsilon\rightarrow\epsilon_{c}$, both $\lambda_{c}$ and $\lambda_{l}$ decrease quickly with $\lambda_{c}<\lambda_{l}$ in a wide range of $\epsilon$. This numerical result agrees with our analysis in Sec.~\ref{sub:Parameters-and-protocol} and reveals that the quantum adiabaticity can be strongly enhanced by choosing appropriate control parameters.

\section{Discussions}
\label{sec:Discussions}

\begin{figure}
\includegraphics[clip,width=8.5cm]{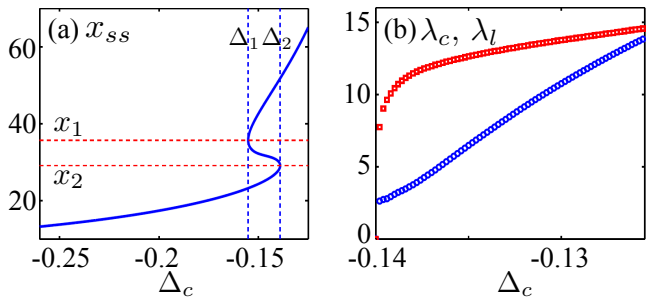}
\caption{(a) $x_{ss}$ vs $\Delta_{c}$ for a TFIM of $N=120$ qubits. Thin dashed lines indicate the detunings $\Delta_{1,2}$ and the displacements $x_{1,2}$ at the bifurcation points. 
(b) $\lambda_{c}$ of the cavity-coupled TFIM (circles) and $\lambda_{l}$ of the linear-ramping model (squares) vs $\Delta_{c}$. 
Here $\epsilon=5$ and other parameters are the same as those in Fig.~\ref{fig1}.}
\label{fig5}
\end{figure}
As an approach that exploits time-dependent ramping to maintain the quantum adiabaticity, the advantage of our approach is that it does not require the spectral knowledge of the quantum many-body system. Instead, this method relies on nonlinear properties that are generic in adiabatic quantum computers to achieve strong enhancement of the quantum adiabaticity. It only requires the knowledge of the operator average $X_{ss}$ and its derivative $X_{ss}^{\prime}$ at $\widetilde{B}_{x}=0$ and $B_{x}$, which can be obtained analytically with perturbation theory, to engineer the system parameters. In contrast, previous works with time-dependent ramping require the spectrum knowledge of the quantum system to design the ramping rate, which is manipulated according to the local energy separation between the ground and the excited states~\cite{SenPRL2008, MondalPRB2009, BarankovPRL2008, RolandCerfPRA2002, HTQuanNJP2010, DoriaPRL2011, RahmaniPRL2011, PowerPRB2013, NWuPRB2015}. Whereas the energy spectrum for questions of interest to adiabatic quantum computing is hard to obtain with classical methods. 

Our approach can be applied to a wide variety of quantum systems with different control parameters and different forms of coupling. In previous sections, we use the driving amplitude as the control parameter. Other parameters can also be utilized as a control knob, such as the detuning $\Delta_{c}$. For a TFIM biased at $\epsilon=5$ with other parameters being the same as those in Fig.~\ref{fig1}, a bistable regime in the detuning exists with two bifurcation points at $\Delta_{1}=-0.155$ and $\Delta_{2}=-0.14$, respectively. The stationary cavity displacement $x_{ss}$ is plotted vs $\Delta_{c}$ in Fig.~\ref{fig5}(a). 
Consider an adiabatic protocol, where the detuning is switched from its initial value to an intermediate value $\Delta_{c}$ at the time $t=0$, and then the system starts evolving towards the stationary state of $\Delta_{c}$. When the effective field is below $\widetilde{B}_{x}=0.8$ during the evolution, the detuning is switched from $\Delta_{c}$ to a final value that corresponds to $\widetilde{B}_{x}=0$ in the stationary state. We conduct numerical simulation under this protocol. In Fig.~\ref{fig5}(b), the ramping rate $\lambda_{c}$ at the critical point is shown together with the linear-ramping rate $\lambda_{l}$ defined in Sec.~\ref{sec:cavityTLS}. Strong enhancement of the quantum adiabaticity in comparison with the linear-ramping model can be observed. 

Furthermore, other forms of coupling can be utilized to implement our approach as well. One example is the photon-number coupling $H_{int}=ga^{\dag}a\sigma_{xi}$. In \cite{TianPRL2010}, it was shown that bistable behavior can be observed with this coupling. Our numerical study finds that the quantum adiabaticity can be enhanced with this interaction.

\section{Conclusions}
\label{sec:conclusions}

To conclude, we studied a universal approach to enhancing the quantum adiabaticity in continuous quantum processes by coupling an adiabatic quantum computer to an ancilla cavity. This approach exploits the nonlinear features that are generic to adiabatic quantum computers to engineer the system parameters and position the gap region between the bifurcation points. The time evolution in the gap region can be significantly slowed down and the quantum adiabaticity can be strongly enhanced. We applied this method to a quantum TLS, a randomly-generated instance of the EC problem, and a TFIM using numerical simulation. The advantage of this method compared with previous methods is that it does not require the spectral knowledge of the adiabatic quantum computer or the engineering of unphysical interactions. This approach can be applied to a vast variety of adiabatic quantum processes when combined with the cavity or circuit QED technology.

\begin{acknowledgments}
This work is supported by the National Science Foundation (USA) under Award Number DMR-0956064, the UC Multicampus-National Lab Collaborative Research and Training under Award No. LFR-17-477237, and the UC Merced Faculty Research Grants 2017. 
\end{acknowledgments}


\begin{thebibliography}{References}
\bibitem{NoteCoupling}The coupling Hamiltonian $H_{int}$ is in the rotating frame of the driving frequency $\omega_{d}$. The interaction in the lab frame is $H_{int}^{(0)}=2g\cos(\omega_{d}t)(\hat{a}+\hat{a}^{+})\sum_{i}\sigma_{xi}$.
\bibitem{NoteWavefunction}The wave function of the TFIM can also be written as $|\psi(t)\rangle=\prod_{k>0}[\alpha_{k}(t)+i\beta_{k}(t)\gamma_{k}^{\dag}\gamma_{-k}^{\dag}]\vert G_{k}\rangle$, where $\gamma_{k}$ ($\gamma_{k}^{\dag}$) is the annihilation (creation) operator of the quasiparticle with the quasimomentum $k$, $\alpha_{k}$ ($\beta_{k}$) is the probability amplitude of the ground (excited) state, and $|G_{k}\rangle$ is the ground state in the subspace of the quasimomenta $\pm k$.
\end{thebibliography}
\end{document}